\documentclass[twocolumn,article,superscriptaddress,longbibliography,10pt]{revtex4-1}
\usepackage{amsmath}
\usepackage{amssymb}
\usepackage{graphicx}
\usepackage{xcolor}

\usepackage{times}
\usepackage{setspace}

\usepackage{multirow}

\usepackage[utf8]{inputenc}
\usepackage{gensymb}

\usepackage{framed}
\usepackage{soul}
\colorlet{shadecolor}{blue!20}

\begin{document}

\title{Structure and flow of low-dimensional water}
\author{Maxim Trushin}
\affiliation{Department of Material Science and Engineering, National University of Singapore, 9 Engineering Drive 1, 117575, Singapore}
\affiliation{Institute for Functional Intelligent Materials, National University of Singapore, 4 Science Drive 2, 117544, Singapore}
\author{Daria V. Andreeva}
\email{daria@nus.edu.sg}
\affiliation{Department of Material Science and Engineering, National University of Singapore, 9 Engineering Drive 1, 117575, Singapore}
\affiliation{Institute for Functional Intelligent Materials, National University of Singapore, 4 Science Drive 2, 117544, Singapore}
\author{Francois M. Peeters}
\affiliation{Department of Physics, University of Antwerp, Groenenborgerlaan 171, B-2020 Antwerp, Belgium}
\affiliation{Departamento de Física, Universidade Federal do Ceará, Fortaleza-CE 60455-760, Brazil}
\author{Kostya S. Novoselov}
\email{kostya@nus.edu.sg}
\affiliation{Department of Material Science and Engineering, National University of Singapore, 9 Engineering Drive 1, 117575, Singapore}
\affiliation{Institute for Functional Intelligent Materials, National University of Singapore, 4 Science Drive 2, 117544, Singapore}

\begin{abstract}    
Water, a subject of human fascination for millennia, is likely the most studied substance on Earth, with an entire scientific field
--- hydrodynamics --- dedicated to understanding water in motion. 
However, when water flows through one-dimensional or two-dimensional channels, its behavior deviates substantially from the principles of hydrodynamics.
This is because reducing the dimensionality of any interacting physical system amplifies interaction effects 
that are beyond the reach of traditional hydrodynamic equations.
In low-dimensional water, hydrogen bonds can become stable enough to arrange water molecules into an ordered state, causing water to behave not only like a liquid but also like a solid in certain respects. In this review, we explore the relationship between water's ordering and its ability to flow in low-dimensional channels, using viscosities of bulk water, vapor, and ice as benchmarks. We also provide a brief overview of the key theoretical approaches available for such analyses and discuss ionic transport,
which is heavily influenced by the molecular structure of water.

\end{abstract}

\maketitle

\section{Introduction}
\label{intro}

The intermolecular interactions in water arise from hydrogen bonds,
which can organize the water molecules into an ordered state.
The history of water structure research has often been a topic of scientific controversy,
beginning with the polywater hypothesis in the late 1960s,
which implied that water molecules could form chains in small capillary tubes \cite{rousseau1970polywater,derjaguin1983polywater}.
By the mid-1970s, polywater was debunked as a misconception, but ironically, molecular water chains were actually discovered three decades later in carbon nanotubes (CNTs) \cite{hummer2001water,majumder2005enhanced,holt2006fast,kofinger2008macroscopically}.
Later on, water monolayers have been found in several layered confinements 
\cite{algara2015square,bampoulis2016coarsening,severin2012reversible,kim2013between,song2014evidence} igniting 
scientific debates about its structure \cite{zhou2015observation,mario2015aa}.
The water molecular chains and monolayers are examples of extreme dimensionality reduction,
which can be seen as one-dimensional (1D) and two-dimensional (2D) water, respectively.
In these low-dimensional forms, water maintains an ordered molecular structure and exhibits flow behavior that differs from its bulk counterpart.
The exploration of water transport at such a small scale
has only emerged in the past 20 years \cite{bocquet2020nanofluidics}, giving rise to the field of nanofluidics
\cite{emmerich2024nanofluidics} and even angstrofluidics \cite{boya2024wonderland}.

\begin{figure}
 \includegraphics[width=\columnwidth]{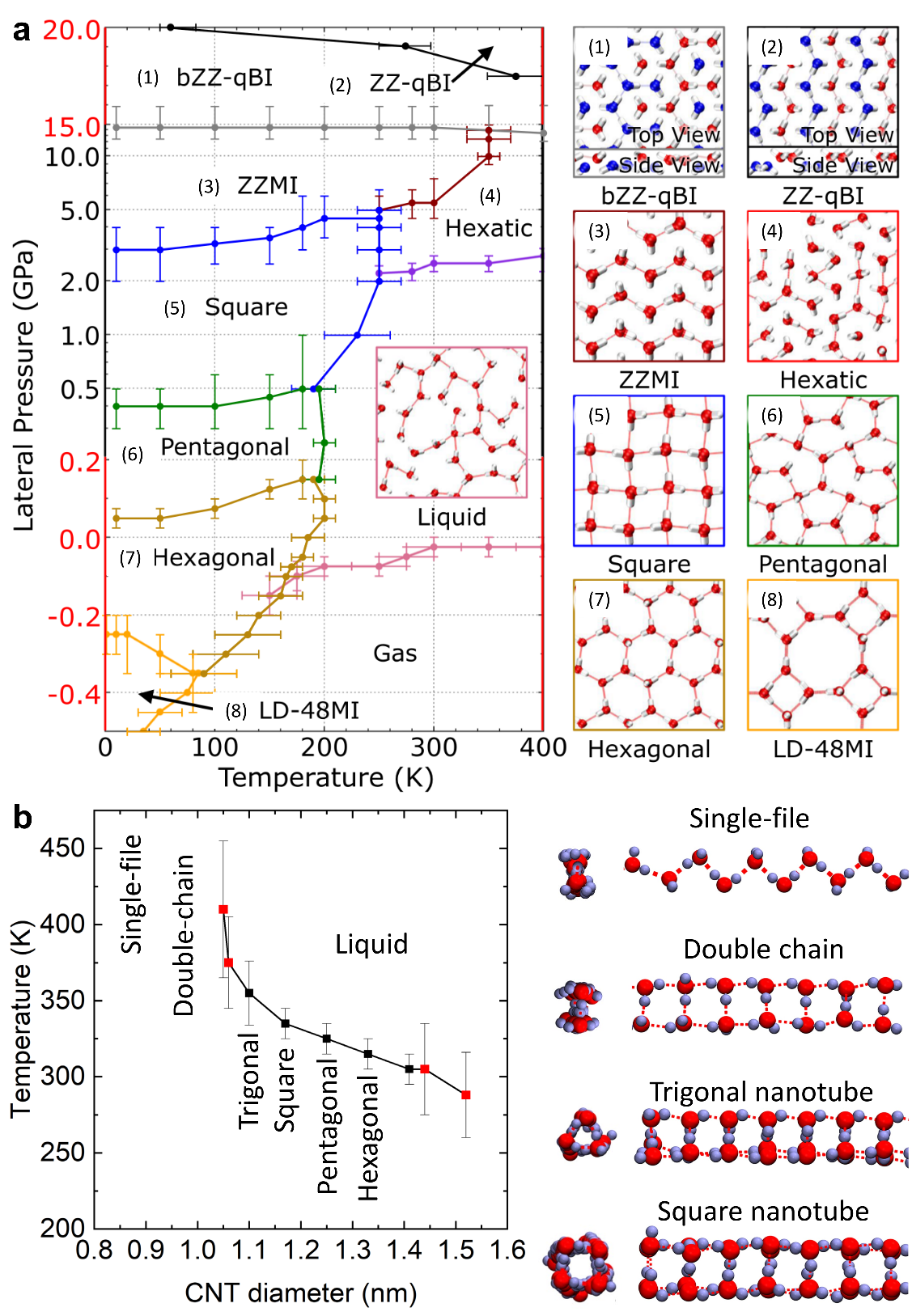}
 \caption{Phase diagrams of confined water in 2D and 1D with the representative molecular structures.
 (a) The phase diagram of monolayer water/ice in nanoconfinement with hydrophobic walls
 determined by means of MD simulations with machine-learning force fields \cite{lin2023temperature}.
 The vertical axis in black and red are in the logarithm and linear scale, respectively,
 bZZ-qBI, ZZ-qBI, ZZMI, LD-48MI stand for branched zigzag quasi-bilayer ice, 
 zigzag quasi-bilayer ice, zigzag monolayer ice, low-density $4\cdot 8^2$ monolayer ice, respectively.
 (b) The solid-liquid phase diagram for water confined in a CNT depending on its diameter. 
 Black and red points are obtained from MD simulations \cite{raju2018phase}
 and deduced from Raman spectroscopy measurements \cite{NatureNano2017}, respectively. 
 Upon lowering temperature water transitions into various $n$-gonal ice structures depending on the diameter of CNT.
 No first-oder transition occurs for single and double chains, see the main text.
 The images are reproduced from \cite{lin2023temperature} and \cite{raju2018phase}.
}
 \label{fig0}
\end{figure}

The static structure of low-dimensional water has been studied for decades \cite{cui2023low}
through molecular dynamics (MD) simulations in 1D \cite{takaiwa2008phase}
and 2D \cite{zangi2003monolayer} confinements, as well as experimentally via X-ray diffraction \cite{maniwa2005ordered},
Raman spectroscopy \cite{NatureNano2017}, transmission electron microscopy \cite{algara2015square},
and neutron scattering \cite{kolesnikov2004anomalously}.
Figure \ref{fig0}a,b shows the typical phase diagrams for low-dimensional water
with the representative structures. 
The most recent machine-learning assisted {\it ab initio} studies \cite{kapil2022first,lin2023temperature}
agree that the 2D liquid phase is adjacent to the hexagonal, pentagonal, square, and hexatic monolayer ices
as well as to the gas phase at low pressure. 
Since the typical van der Waals pressure is 1--2 GPa \cite{vasu2016van}, the default water phase state
in the 2D limit is expected to be liquid or hexatic at room temperature (Fig. \ref{fig0}a).
The hexatic phase implies a certain structure but no long-range order, making 2D water neither liquid nor crystalline.
However, in contrast to the phase diagram of bulk water, where transitions are primarily influenced by pressure and temperature, the state of nanoconfined water is also sensitive to factors like 
the size of the confinement \cite{giovambattista2009phase} and the properties of the confining material (e.g., hydrophobic/hydrophilic  \cite{giovambattista2006effect}, polar/non-polar \cite{negi2022edge}). For example, a monolayer of water can transition from a liquid to a crystalline phase and then back to a liquid state with even a tiny increase in the distance between the confining plates (less than 1\AA), all while maintaining constant lateral pressure and room temperature ($\sim 300$ K) \cite{zangi2003monolayer}.
Similarly, the liquid-solid transition temperature for water in CNTs is determined by the diameter of the CNT,
increasing by tens of K when the diameter reduces by a few \AA. 
Upon crystallization, water adopts various structures, depending on the diameter (Fig. \ref{fig0}b).
However, in the 1D limit, no first-order transition occurs, and water consistently forms molecular chains, regardless of temperature \cite{raju2018phase}.

The structure of water dramatically affects its kinetic properties, such as flow. 
However, a key challenge lies in the quantitative interpretation of low-dimensional water flow measurements \cite{majumder2005enhanced,2Dslit2016}
using traditional hydrodynamic parameters like slip length and viscosity \cite{bocquet2010nanofluidics}.
Slip length, written as $l_s=\eta/\lambda$, refers to the ratio of internal friction (described by viscosity $\eta$)
to the surface friction coefficient, $\lambda$, which reflects the interactions between water and the confining material.
In low-dimensional water, though, it's difficult to distinguish between bulk and surface behaviors, making the standard definitions of internal and surface frictions unclear.
Indeed, it was recognized a long time ago \cite{hanasaki2006flow}
that both quantities strongly depend on confinement size, confining material, and experimental settings
\cite{2Dslit2016,APL2018peeters,keerthi2021water},
hence, they naturally change from sample to sample complicating the analysis
of MD simulations and experimental data \cite{thomas2008reassessing,neek2016commensurability}.
To give an example,
the remarkable ability of carbon nanochannels to enable nearly frictionless transport remains a debated phenomenon, despite recent advances in both theoretical \cite{kavokine2022fluctuation} and experimental \cite{wu2024probing} research.
Nevertheless, despite the conceptual challenges \cite{faucher2019critical},
there has been no shortage of efforts to observe or simulate water transport 
at the nanoscale \cite{aluru2023fluids,gao2017nanofluidics} and beyond \cite{shen2021artificial},
as well as to explore the potential use of nanoconfinement in water filtration devices  \cite{werber2016materials}.

In this Review, we aim to connect the fundamental issue of low-dimensional water transport
with the molecular-level structural changes in water flow. 
Section \ref{viscosity} begins by comparing 1D and 2D water to other forms of water, using viscosity as a key characteristic.
Viscosity serves as a key measure of internal friction, which is closely tied to the molecular structure of water.
Since low-dimensional water flow requires strong confinement, Section \ref{slip} explores the role of boundary conditions and external friction in low-dimensional limits. 
The relationship between water structure and flow leads to several filtration mechanisms for ionic selectivity, 
which we discuss in Section \ref{selectivity}. Finally, we conclude with a perspective on future developments in Section \ref{outro}.

\section{Low-dimensional water in motion}
\label{viscosity}

Both liquids and solids are composed of molecules held together by intermolecular bonds.
The apparent difference between the two is that liquids can flow, while solids cannot.
However, this distinction is more quantitative than qualitative, at least when it comes to water.
Water, along with over a dozen different ice phases, is made up of the same molecules bound by the same forces,
and it can flow regardless of its phase.
This concept has likely been known to humans since ancient times, who observed glaciers flowing, even before modern scientists were able to study the viscosity of water ice and other solids in controlled environments \cite{glass_viscosity_1996}. 
In this context, we explore the correlations between structural transitions in low-dimensional water and the accompanying changes in viscosity.

\begin{shaded}
{\bf Box 1. Viscosity evaluation.}

\vspace{0.5cm}

There are four principal methods for determining the viscosity of water ($\eta$) in Table \ref{tab1}:
\begin{itemize}
    \item  Measuring the water flow rate \cite{2Dslit2016} and inferring $\eta$ based on a chosen continuum model \cite{APL2018peeters}.
    \item  Running multiple MD simulations to compute the water flow rate and again deducing  $\eta$ using a continuum model \cite{trushin2023two}.
    \item  Extracting $\eta$ directly from MD simulation data via the Green–Kubo linear response relation \cite{bocquet1994hydrodynamic} --- this can be referred to as the ‘physical’ method.
    \item Conducting MD simulations and determining $\eta$ using the Eyring method, which treats viscous 
    flow as a chemical reaction \cite{eyring1936viscosity} --- this can be referred to as the ‘chemical’ method
\end{itemize}

The first and second methods share an obvious limitation: the validity of any continuum model is not assured at length scales below 2 nm \cite{bocquet2010nanofluidics}. Meanwhile, MD simulations are highly dependent on the chosen force-field model, and no single model can perfectly replicate all experimental values.
For example TIP3P works rather well for the thermodynamical properties but overestimates diffusivity while SPC/E does just vice versa.
This necessitates careful selection of a water model tailored to the specific phenomena under investigation \cite{kadaoluwa2021systematic}.
The viscosity values for low-dimensional water summarized in Table 1 are mostly deduced from MD simulations using different 
force-field models.

\end{shaded}

\begin{table*}
\begin{tabular}{|l|c|c|l|c|}
\hline
State of water & $\eta$ (Pa$\cdot$s) & $d$ (nm) & Comments and references & Structure \\
\hline
Bulk vapor & $ 10^{-5}$  & $\infty$ & measured at 310 K \cite{hellmann2015viscosity} & 
\includegraphics[width=0.09\textwidth]{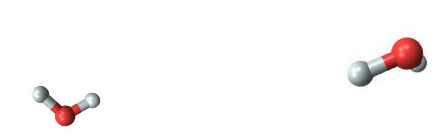} \\
\hline
Confined by a CNT & $\sim 0.2 \times 10^{-4}$  & $0.7$ & continuum model estimation \cite{myers2011slip} & 
\multirow{2}{*}{\includegraphics[width=0.1\textwidth]{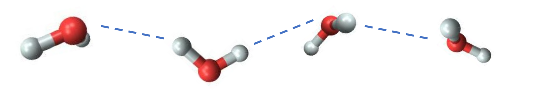}}\\
& & & using experimental data \cite{whitby2008enhanced} &  \\
Confined by a CNT & $ 0.4 - 4 (\times 10^{-4})$  & $0.81 - 5.42$ & TIP3P MD simulations at 298 K \cite{babu2011role} &  
\multirow{3}{*}{\includegraphics[width=0.1\textwidth]{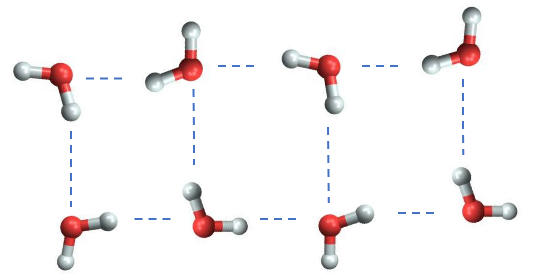}} \\
Confined by a CNT & $ 1.7\times 10^{-4}$  & $0.81$ & TIP4P-EW MD simulations at 298 K \cite{zhang2011prediction} & \\
Confined by a CNT & $ 6.5 - 8 (\times 10^{-4})$  & $1.66 - 4.99$ & TIP5P MD simulations at 298 K \cite{thomas2008reassessing} &  \\
\hline
Bulk water & $  0.28 - 1.79  (\times 10^{-3})$  & $\infty$ & measured from 0 to 100\degree C  & 
\multirow{4}{*}{\includegraphics[width=0.09\textwidth]{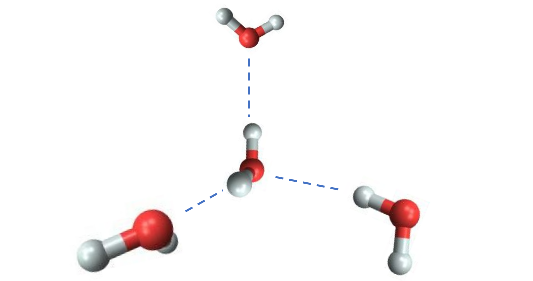}}  \\
 & & & at ambient pressure \cite{korson1969viscosity} & \\
Bulk water & $ 0.56 - 3.59  (\times 10^{-3})$  & $\infty$ & TIP4P/Ice MD simulations \cite{baran2023self}  & \\
Bulk water & $ 1 - 3 (\times 10^{-3})$  & $\infty$ & measured at 21 \degree C and a few GPa \cite{abramson2007viscosity} & \\
\hline
Supercooled water &  $ 1.8 - 5.5 (\times 10^{-3})$  & $\infty$ &  measured from 0 to $-24$\degree C & 
\multirow{6}{*}{\includegraphics[width=0.09\textwidth]{bulk-water.eps}} \\
 & & &  at ambient pressure \cite{hallett1963temperature} &  \\
Supercooled water &  $ 4 \times 10^{-3}$  & $\infty$ &  measured at 244 K  &  \\
 & & & pressure up to 0.3 GPa \cite{singh2017pressure} & \\
Supercooled water&  $ 1 - 15 (\times 10^{-3})$  & $\infty$ &  measured down to 239 K
\cite{dehaoui2015viscosity} & \\
Supercooled water&  $ 1 - 30 (\times 10^{-3})$  & $\infty$ &  TIP4P/2005f MD simulations \cite{guillaud2017decoupling} &  \\
\hline
Confined by charged GO &  $3 \times 10^{-3}$ &  $0.7$ & SPC/E MD simulations at 300 K \cite{kalashami2018slippage} &
\multirow{5}{*}{\includegraphics[width=0.15\textwidth]{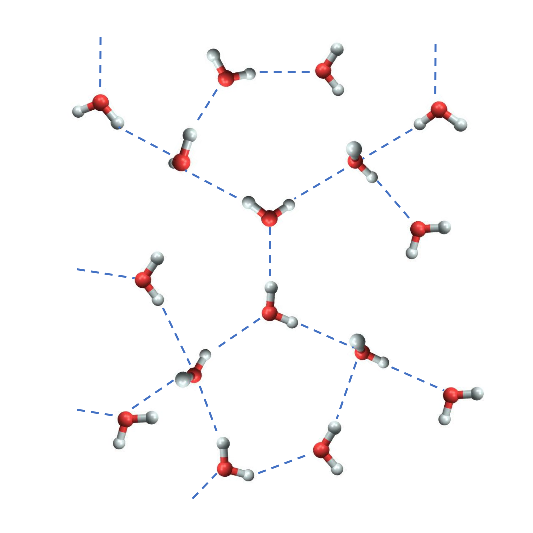}}\\
Confined by graphene &  $6.5 \times 10^{-3}$ &  $0.6$ & SPC/E MD simulations &  \\
 & & & fitted by a continuum model \cite{trushin2023two} & \\
QLL at ice/solid interface & $ 8-10 (\times 10^{-3})$  & $0.8-1.2$ & TIP4P/Ice MD simulations  &   \\
 & & & at 1 atm and 262 K \cite{baran2022ice} & \\
Confined by neutral GO &  $0.02$ &  $0.7$ & SPC/E MD simulations at 300 K \cite{kalashami2018slippage} & \\
QLL at ice/solid interface & $ 0.03$  & $0.82$ & TIP4P/Ice MD simulations at 240 K & 
\multirow{4}{*}{\includegraphics[width=0.15\textwidth]{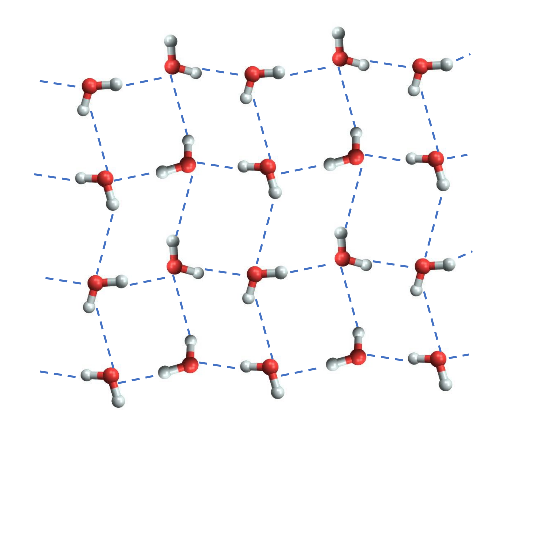}} \\
& & & low sliding speed \cite{zhao2022new} &  \\
Confined by graphene &  $0.03$ &  $0.8$ & flow rate measurements \cite{2Dslit2016}  &  \\
 & & & fitted by a continuum model \cite{APL2018peeters} & \\
QLL at ice/water interface & $ 0.035 $  & $0.68$ & TIP4P/Ice MD simulations 
at 255 K \cite{louden2018ice} &  \\
Confined by h-BN &  $0.035$ &  $0.7$ & SPC/E MD simulations &
\multirow{4}{*}{\includegraphics[width=0.15\textwidth]{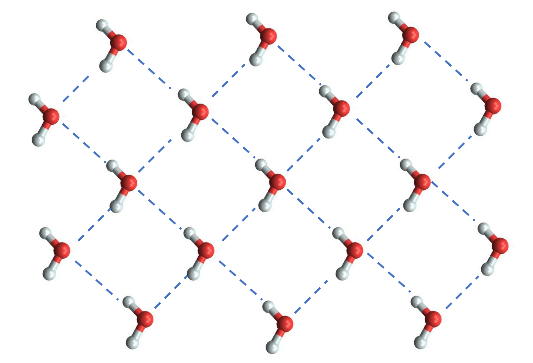}}  \\
& & &  fitted by a continuum model \cite{trushin2023two} & \\
Confined by mica &  $0.042$ &  $0.92$ & TIP4P MD simulations at 298 K \cite{leng2005fluidity} &  \\
Confined by graphene &  $0.0003-0.3$ & $0.75-0.9$ & ReaxFF MD simulations  &  \\
& & & at a constant density \cite{neek2016commensurability} & \\
\hline
Ice VI & $0.84 \times 10^{12}$  & $\infty$ & measured at 260 K and 6 kbar \cite{sotin1987viscosity} & \multirow{3}{*}{\includegraphics[width=0.15\textwidth]{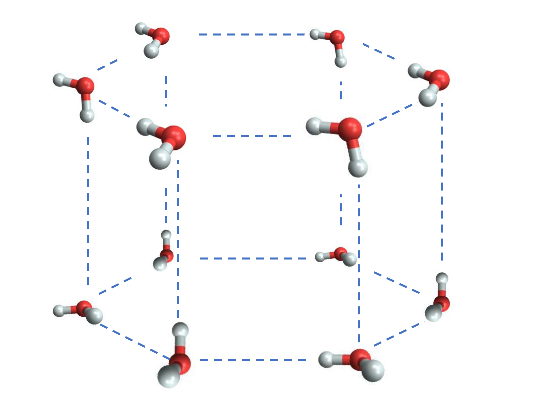}}\\
Ice V & $45 \times 10^{12}$ & $\infty$ & measured at 260 K and 6 kbar  \cite{sotin1987viscosity} &  \\
Glacier ice  & $ 0.3 - 29.2 \times 10^{12}$  & $\infty$ & observations of Swiss glaciers \cite{deeley1908viscosity} &  \\
Ice XI & $>10^{12}$ & $\infty$ & expected below 72 K and 0.07 GPa \cite{yen2015proton} &  \\
& & &  & \\
\hline
\end{tabular}
\caption{Shear viscosity of water and its schematic molecular structure under various conditions. 
The lines are arranged from the lowest to the highest viscosity.
The confinement size $d$ stands for either nanotube diameter (1D confinement) or channel height (2D confinement).
The representative structures demonstrate the correlation between the average number of hydrogen bonds and internal friction.
Note that this is a very simplified picture because the viscosity is also determined by the strength, orientation, and lifetime of hydrogen bonds.
These parameters are not shown in the table but they are distinguishing factors between supercooled water and water at ambient conditions, as well as between 2D and 3D ice phases. 
}
\label{tab1}
\end{table*}

Table \ref{tab1} shows a correlation between the structure and viscosity of water. The rows are arranged from the lowest to the highest viscosity. The rightmost column provides a schematic representation of typical water structures corresponding to several neighbouring rows. It is important to note that the hydrogen bonds, indicated by dashed lines, can vary in strength and lifetime depending on the phase. 
Stronger and more stable bonds generate greater internal friction, which in turn increases the overall viscosity.
The table illustrates this trend by showing a higher average number of bonds per water molecule as one moves from the upper to the lower rows.

Starting with the topmost row, one can see that bulk water vapor has the lowest viscosity coefficient, which correlates with the complete absence of any structure and active hydrogen bonds.
Water confined in 1D forms dipole chains \cite{kofinger2008macroscopically}, where each water molecule
has one or two neighbours linked by the dipole-dipole interactions.
As a consequence, viscosity of 1D water is greater than that of water vapor but does not reach the values
typical for bulk water. The reason for that is the larger average number ($\sim 3$) of hydrogen bonds   formed by each molecule when the confinement is lifted \cite{muthachikavil2021distinguishing}.

\begin{shaded}
{\bf Box 2. Water models.}

\vspace{0.5cm}

The development of intermolecular potentials for water started in the 1960s \cite{barker1969structure}
and is still an important topic of research.  An overview of some of those `water' models can be found on the website of
M. Chaplin \cite{water-online}.
One of the early and still most popular force field interatomic potentials is TIPnP/year, 
where `n' refers to the number of points in the potential, and the `year' of the improved version of TIPnP. 
These are rigid models, i.e. the H--O--H angle and the O--H distance within a water molecule are fixed and not polarizable; see the image below for TIP3P. 

\vspace{0.5cm}
\includegraphics[width=0.9\columnwidth]{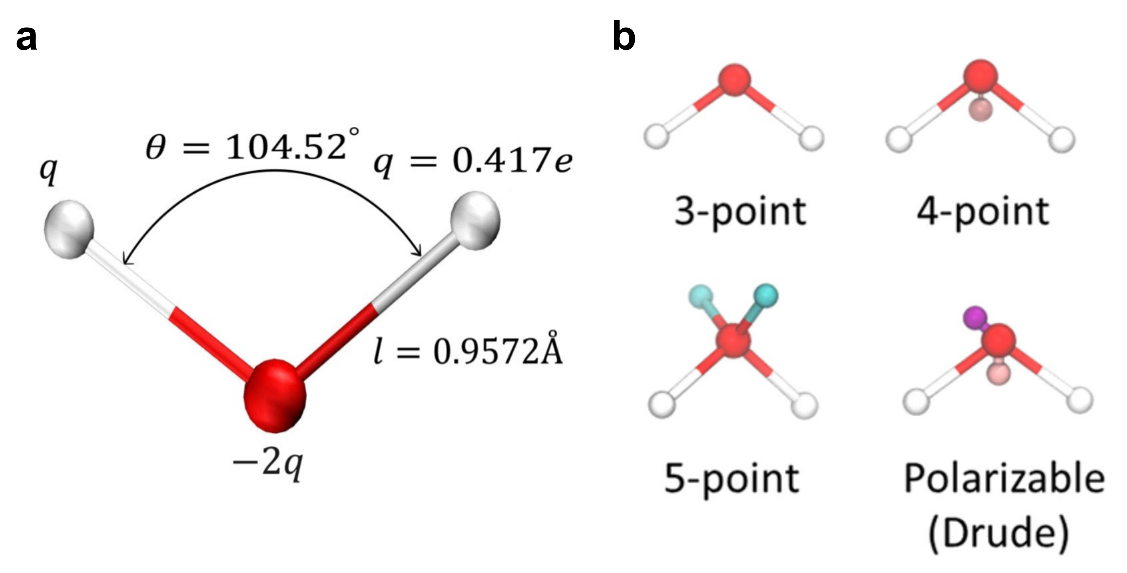}
 
 (a) The 3-point water model TIP3P. Image reproduced from \cite{izadi2016accuracy}.
 (b) Sketches of the n-site water models. The offset partial charge on oxygen in 4-point
models is colored pink. The lone pairs in 5-point models are colored cyan.
The Drude oscillator in the polarizable model is colored purple. Images reproduced from \cite{kadaoluwa2021systematic}.
\vspace{0.5cm}

In four-site rigid models, the charge that is usually assigned to the oxygen atom is displaced from the centre of this atom, thus requiring one additional site to completely describe a water molecule.
The simple point charge (SPC) models assume an ideal tetrahedral shape (H--O--H angle of 109.47°) instead of the observed angle of 104.5°. Over the years several extensions of the rigid models have been proposed to include, for example, molecular flexibility, non-pairwise interactions and polarization effects. The 30 commonly used n-point ($n=3,4,5$) and polarizable water models are reviewed in \cite{kadaoluwa2021systematic}. 
An important advance has been the development of ReaxFF (reactive force field) that employs a bond-order formalism in conjunction with a fluctuating charge description of polarization \cite{senftle2016reaxff}.
It allows for molecular flexibility, polarization, and chemical reactivity (bond formation and breaking).
Another issue is the water channel wall interaction. A repository that provides useful interatomic potentials (force fields) for such an interaction can be found at the website \cite{water-conf-online}.

\end{shaded}

There is experimental \cite{modig2003temperature} and theoretical \cite{muthachikavil2021distinguishing} evidence 
that the strength of hydrogen bonds depends on pressure and temperature.
When water is supercooled, the average number of active hydrogen bonds may increase, 
which manifests in a larger viscosity coefficient, as seen in Table \ref{tab1},
where an extra bond is added to the standard tetragonal representation of a water molecular cluster.
Stronger bonds lead to greater interactions between water molecules and increased internal friction in supercooled water. However, it is important to note that hydrogen bonds are feeble in water (with lifetimes on the order of 1 ps \cite{martiniano2013insights}), and the diagrams in the rightmost column of the table do not depict static structures.
The only exception is bulk water ice, shown in the lowest rows of the table, where the hydrogen bonds are fully saturated in all directions, forming a true crystalline structure with a very high shear viscosity.
Bulk ice is represented by the structure of ice-XI, a hydrogen-ordered form of ice with the lowest internal energy \cite{biggs2024proposing}, which is thought to have the most stable and strongest intermolecular bonds \cite{zivk2025water}.

Water confined in 2D can acquire very different structures \cite{zhao2014highly} depending not only on 
pressure and temperature but also on confining material, and so does viscosity.
MD simulations of water monolayer under high pressure of about 2-3 GPa indicate hexatic phase \cite{lin2023temperature},
which can be seen as the second least structured phase right below the supercooled liquid state in Table \ref{tab1}.
It is expected to occur in a tight confinement formed by the van der Waals forces and demonstrate the lowest viscosity. 
The square or puckered rhombic phase are obviously more structured and should demonstrate somewhat higher viscosity,
as the in-plane shear requires breaking more hydrogen bonds.
However, there are 2D water structures with even higher viscosity due to a finite net polarization
induced by polar confining materials, like h-BN or mica. One of such structures is the rhombic one
shown in Table \ref{tab1}, but one should have in mind that the realistic structured water flow
involves domains of different orientations \cite{trushin2023two}.

\subsection{Viscosity of 1D water}

1D water flow was first observed in carbon nanotubes (CNTs) about two decades ago \cite{majumder2005enhanced,holt2006fast}, and the flow rates exceeded
by far the values expected from naive continuum hydrodynamics models.
The water flow enhancement has been initially explained by extremely long slip lengths (longer than a micron) 
due to strongly reduced friction between water and carbon.
The explanation has raised controversy \cite{qin2011measurement}.
The slip length is supposed to be a material parameter independent of the CNT diameter.
However, the slip lengths have been found to vary by orders of magnitude for CNTs of diameters 0.81--10 nm \cite{kannam2013fast}.
On the other hand, the extremely high water permeability of CNTs is a well-established fact confirmed 
by subsequent measurements \cite{majumder2011mass,tunuguntla2017enhanced} and MD simulations \cite{chen2008nanoscale,ye2011size}.
Careful analysis of MD simulation data using
the Green--Kubo relation \cite{thomas2008reassessing} and Eyring equation \cite{zhang2011prediction,babu2011role}
concluded that water viscosity decreases substantially when confined in CNTs of less than 1.5 nm in diameter.
As a consequence, the slip length can be taken within a reasonable range to fit the data even though
the friction between water and CNT walls is indeed reduced \cite{myers2011slip}.

In general, the water viscosity drops by at least one order of magnitude (down to $10^{-4}$ Pa$\cdot$s) 
when confined by a CNT of less than 1 nm in diameter \cite{thomas2008reassessing,zhang2011prediction,babu2011role,ye2011size,chen2008nanoscale}.
The only exception is the older result coming from MD simulations performed at a constant density of the
confined water suggesting a slight increase of water viscosity in narrower CNTs \cite{liu2005fluid}.
Having in mind various water structures with different densities found inside CNTs a few years later 
\cite{takaiwa2008phase,maniwa2005ordered,kofinger2008macroscopically} the constant density assumption does not look feasible.
Table \ref{tab1} suggests that water viscosity in narrow CNTs approaches the viscosity of water vapor
\cite{teske2005viscosity,hellmann2015viscosity}
rather than liquid water \cite{korson1969viscosity}.
Such behavior can be justified by the domination of the single-file transport regime 
when water molecules are ordered along the central line, like train cars pulling through a tunnel, see
the rightmost column in Table \ref{tab1}.
The shear viscosity is reduced in such an ordered flow and may even vanish in a strictly 1D limit.
In such a limit we are left only with a residual (possibly dilational) viscosity 
as a single parameter of internal friction \cite{jaeger2018bulk}.
The dilational viscosity does not exist in incompressible liquids and is often disregarded 
for bulk water but may play an important role in water confined at nanoscale \cite{trushin2023two}.

\subsection{Viscosity of 2D water}

While 1D confinement reduces water viscosity, 2D confinement increases it. However, this effect is not universal and varies depending on the confinement size and the properties of the confining material \cite{2Dslit2016,gopinadhan2019complete,keerthi2021water}. 
The only exception is Ref. \cite{zaragoza2019molecular}
that predicts nanoconfined water viscosity lower than the bulk one regardless of dimensionality 
but the viscosity has been found two orders of magnitude lower in a 1D confinement than the 2D limit,
in agreement with the previous studies.
Similar to 1D water, monolayer water also acquires a certain structure \cite{kapil2022first,lin2023temperature} 
and therefore sometimes referred to as 2D ice \cite{zangi2003monolayer,corsetti2016structural}.
As a result, in-plane water flow occurs in an ordered manner; however, unlike in the 1D limit, there is substantial internal friction between neighboring molecular trains.
This internal friction is stronger than in bulk water, leading to an increase in shear viscosity by an order of magnitude (see Table \ref{tab1}).
Intuitively, one might expect the viscosity of 2D ice to be comparable to that of naturally occurring ice. 
However, the viscosity of crystalline ice is at least 13 orders of magnitude higher than that of a water monolayer. Thus, the internal friction of 2D water falls between that of bulk liquid water and solid ice, though it is much closer to the former than the latter.

Surprisingly, water supercooled to 240 K exhibits a viscosity comparable to that of a water monolayer at 300 K. Since supercooled water is amorphous, its relatively high viscosity cannot be attributed to crystallisation. This suggests that viscosity and crystallisation in water confined within a 2D channel may not necessarily be linked. The striking similarity between confined and supercooled water has led many researchers to explore the possibility of a universal model that applies to both, see Ref. \cite{cerveny2016confined} for review.
However, the problem remains highly complex, as supercooled water and ice can coexist within nanoconfinement \cite{schiller2024ice}.

Interestingly, 2D water confinement and flow occur naturally on the surface of ice sliding on another solid \cite{baran2022ice}. 
The thickness of such a quasi-liquid layer (QLL) and its viscosity both depend on sliding speed \cite{zhao2022new},
temperature \cite{bluhm2002premelting}, and even ice crystal facets \cite{louden2018ice} 
with the typical values are less than 1 nm for confinement size
and of the order of 0.01 Pa$\cdot$s for viscosity.
This very thin outer QLL is known to make the surface of ice slippery \cite{liefferink2021friction}.
Besides its obvious importance in everyday life, QLL also plays a fundamental role in understanding 2D water transport. 
Almost identical viscosity values of QLL and water monolayer suggest that water acquires the same state in both cases, which seems to be neither crystallized ice nor liquid water.

We therefore refrain from referring to a water monolayer as ‘ice,’ as 2D water may or may not crystallise depending on the nature of its confinement. In the 1D limit, water is certainly structured into a linear arrangement, yet it does not exhibit the properties of ice. Thus, the ordering of the water structure is not necessarily equivalent to crystallisation and may instead represent a continuous process that does not involve a first-order transition \cite{raju2018phase,ma2022continuous,han2010phase}. However, the structural organisation of water substantially influences internal friction, much like in solids. These seemingly contradictory characteristics highlight the solid-liquid duality of low-dimensional water.

\section{Hydrodynamic equations in low-dimensional limits}
\label{slip}

\begin{figure}
 \includegraphics[width=\columnwidth]{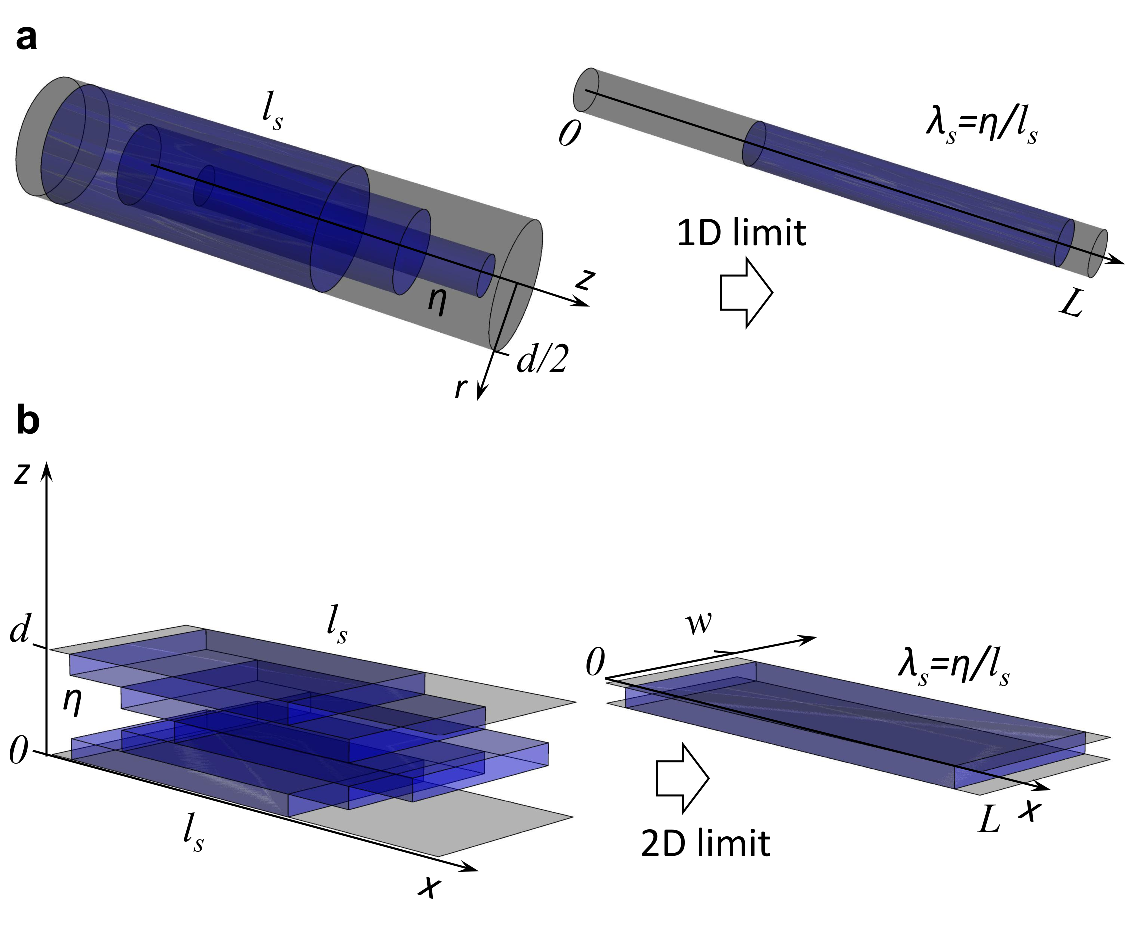}
 \caption{Continuum description of low dimensional water.
Laminar Poisseuille flow in the tube-like (a) and box-like (b) geometries 
with 1D and 2D limits shown on the right. In the 1D and 2D limits, the slip length and viscosity collapse
to a single quantity --- the friction coefficient, $\lambda=\eta/l_s$.
}
 \label{fig1}
\end{figure}

Since the internal volume of 1D and 2D water is negligible, the boundary conditions are crucial for the transport characteristics.
The boundary conditions naturally occur in continuum models based on the Navier-Stokes equation, which in the stationary limit can be written as \cite{landau-hydrodynamics}
\begin{equation}
\label{N-S}
  \rho \left(\vec{v} \cdot \nabla  \right) \vec{v}  =
 -\mathrm{grad}\, p
 + \eta \Delta \vec{v}  +\left(\zeta + \frac{\eta}{3}\right) \mathrm{grad}\,\mathrm{div}\, \vec{v}, 
\end{equation}
where $\vec{v}$ and $\rho$ are the flow velocity and density, respectively,
$p$ is the pressure, $\eta$, $\zeta$ are the viscosity coefficients.
In the Poisseuille regime, a laminar flow is assumed, FIG. \ref{fig1}(a,b),
and the fluid is supposed to be incompressible ($\mathrm{div}\, \vec{v}=0$).
In a tube-like geometry with the pressure difference $\Delta p$ applied along the $z$-axis
the boundary conditions are imposed on a $z$-component of the velocity being a function of the radial coordinate as
$v_z(r)|_{r=d/2}=v_s$, $\partial_r v_z(r)|_{r=d/2}=-v_s/l_s$,
where $v_s>0$ and $l_s$ are the slip velocity and length, respectively.
The mass flow rate then reads
\begin{equation}
 Q_\mathrm{tube}=\frac{\pi\rho\Delta p}{8\eta L}\left(\frac{d}{2}\right)^4\left(1+\frac{8l_s}{d}\right).
 \label{tube}
\end{equation}
In a box-like geometry, the pressure difference applies along the $x$-axis, and
the boundary conditions are imposed on a $x$-component of the velocity being a function of the out-of-plane coordinate $z$ as
$v_x(z)|_{z=0,d}=v_s$, $\partial_z v_x(z)|_{z=0,d}=\pm v_s/l_s$.
The mass flow rate then reads
\begin{equation}
 Q_\mathrm{box}=\frac{\rho w\Delta p }{12\eta L}d^3\left(1+\frac{6l_s}{d}\right).
 \label{plane}
\end{equation}

Equations (\ref{tube}) and (\ref{plane}) have been used to analyze the seminal water flow rate measurements 
in 1D \cite{holt2006fast} and 2D \cite{keerthi2021water} nanochannels, respectively.
However, the concept of Poisseuille flow is doubtful at nanoscale \cite{kavokine2021fluids}.
The incompressibility assumption implies that the nanoconfined water density remains the same as in the bulk but
many MD simulations \cite{zhao2014highly,gao2018phase,qiu2015water,mario2015aa,fernandez2016electric,zubeltzu2017simulations}, 
including first-principles \cite{chen2016two,corsetti2016structural,corsetti2016enhanced} 
and machine-learning \cite{ghorbanfekr2020insights,kapil2022first,lin2023temperature} assisted studies,
show that this is not true.
The simple concentric laminar flow shown in FIG. \ref{fig1}a also contradicts
MD simulations in nanotubes demonstrating a helix flow \cite{liu2005fluid}.
The same applies to FIG. \ref{fig1}b, where the laminar flow resembling a deck of cards
does not take into account the in-plane shear between the neighbouring water molecular trains.
In general, the viscosity is a tensor, and it is strongly anisotropic at nanoscale \cite{liu2005fluid,kalashami2018slippage}, hence,
equations (\ref{tube}) and (\ref{plane}) may fail even for that simple reason.
On the other hand, one can average all tensor components and estimate
the overall effect of internal friction by a single number, $\eta$.

To mitigate the flaws of equations (\ref{tube}) and (\ref{plane}) applied at nanoscale
one can assume that either $\rho$ or $\eta$ (or both) are functions of $d$.
This is the usual starting point of any continuum model beyond the Poisseuille approach.
In its simplest version, viscosity is assumed to be dependent
on the nanotube diameter \cite{thomas2010pressure} or a radial coordinate within the nanotube's volume \cite{myers2011slip,calabro2013modelling}.
In that way, one can reproduce a theoretical enhancement of the flow rate observed previously \cite{holt2006fast,whitby2008enhanced}.
Assuming $\rho$ and $\eta$ to follow $1+\gamma \exp(-d/\delta)$ dependence on $d$ 
(with $\gamma$ and $\delta$ being parameters determined from the MD simulations)
the continuum model \cite{APL2018peeters} was able to reproduce the mass rate measured in 
graphene nanochannels of different heights \cite{2Dslit2016}.
The condition of incompressibility has been lifted in another continuum model \cite{trushin2023two}
to demonstrate the non-linear behavior of 2D water flow on driving pressure.
The viscosity parameters obtained from MD simulations are found to be
indeed strongly dependent on confining materials and confinement size \cite{trushin2023two}.

In continuum models, the slip length $l_s$ is usually assumed to be a material parameter computed from MD simulations for CNTs \cite{kotsalis2004multiphase,kannam2013fast}, 
graphene \cite{kumar2012slip}, boron nitride \cite{tocci2014friction}, 
graphite \cite{ramos2016hydrodynamic}, or GO \cite{PRE2014sliplength}.
However, the origin of high slippage in CNTs \cite{secchi2016massive}
and its strong dependence on curvature \cite{thomas2008reassessing,secchi2016massive} remains puzzling. 
The slip lengths determined from MD simulation studies differ by 3 orders of magnitude, varying between nm and $\mu$m for water flow through CNTs with diameters ranging from 0.81 to 7 nm \cite{kannam2013fast}.
Contrary, the slip length did not show any obvious dependence on the channel height
in graphene confinements  \cite{xie2018fast}, 
as it is expected for a material parameter, however, the smallest channel height 
was well above 1 nm so that water certainly was not 2D, see also Refs. \cite{faucher2019critical,sam2021fast} for discussions.

The key conceptual challenge is that the definition of slip length requires a finite volume, as it is determined by linearly extrapolating the fluid velocity profiles, $v_z(r)$ or $v_x(z)$,
beyond the channel walls until the velocity reaches zero.
However, neither 1D nor 2D water has a well-defined volume 
since laminar flow cannot be meaningfully fractionalized below the scale of a single molecule.
Theoretically, the slip length $l_s$ can be eliminated by taking
1D and 2D limits ($d/l_s\to 0$) in equations (\ref{tube}) and (\ref{plane}) that
results in  
$Q_\mathrm{1D}=\pi\rho\Delta p d^3/(16L\lambda)$ and 
$Q_\mathrm{2D}=\rho w \Delta p d^2/(2 L\lambda)$, respectively.
In this limit, the internal and external friction parameters $\eta$ and $l_s$
merge into a single friction coefficient $\lambda=\eta/l_s$.
As a result, in low-dimensional water, it becomes impossible to distinguish between surface and bulk contributions to overall friction.
The friction coefficient was found for water
via MD simulations in graphene \cite{PCCP2017diao} and 
in CNTs \cite{falk2010molecular} to be of the order of $10^4$ Ns/m$^3$,
that is about an order of magnitude lower than on boron nitride \cite{tocci2014friction}.
Such a huge difference could be related to the polar structure of boron nitride \cite{nigues2014ultrahigh}
or electronic ``lubrication'' in graphene \cite{kavokine2022fluctuation} 
and remains a debatable question.

\section{Ionic selectivity mechanisms in low-dimensional water}
\label{selectivity}

\begin{figure}
 \includegraphics[width=\columnwidth]{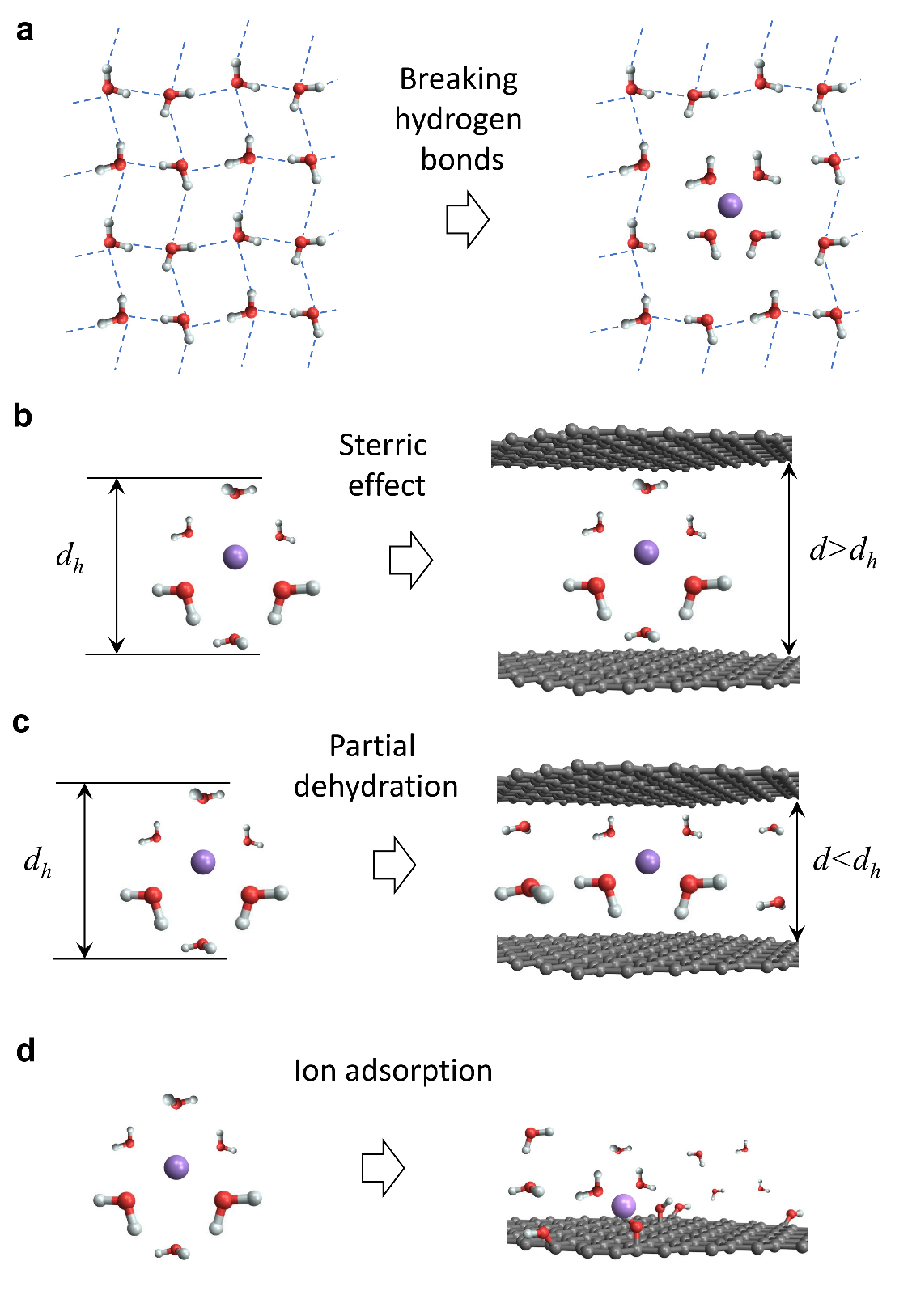}
 \caption{Mechanisms of ionic selectivity in low-dimensional water flow.
 (a) When an ion enters a cluster of low-dimensional water, nearby water molecules must shift, breaking hydrogen bonds and reorienting to form a hydration shell. The associated energy cost, dependent on the water structure and ion species, serves as a selection criterion. 
(b) The hydrated ion diameter, $d_h$, provides a straightforward selectivity mechanism when compared with the confinement size, $d$.
(c) If $d<d_h$, ion entry requires partial dehydration, incurring an energy penalty that can act as an additional filtering mechanism. 
(d) Coulomb interactions between the ion and a charged functional group can further influence ion selectivity by preferentially excluding certain species.}
 \label{fig2}
\end{figure}

Natural water is seldom pure, and the translocation of various solubles often accompanies its flow.
When salts dissolve in water the ions acquire a hydration shell of water molecules \cite{richards2012importance}.
The charge-dipole interactions between polar water molecules and an ion mostly determine the size and structure of the shell.
In addition, mutual dipole-dipole interactions between neighboring water molecules may also play a role.
Different ions form hydration shells of various shapes (octahedral, tetrahedral, etc.) but the size is often characterized by an average hydrated diameter ($d_h\sim 0.6-0.9$ nm) \cite{babu1999theory}. Another important characteristic of hydrated ions is the free energy of hydration or simply dehydration energy taken with an opposite sign \cite{tansel2012significance}. The dehydration energy, $W_h$, shows how much work has to be done to disband the hydration shell completely. The dehydration energy of smaller water clusters follows an approximately linear relation to the number of water molecules remaining in the shell 
\cite{zwolak2009quantized}. Hence,  we can describe the partial dehydration energy as $n_{dh}W_h/n_h$, where
$n_h$ is the total number of water molecules in the shell, and $n_{dh}$ is the number of water molecules removed from the shell. In this section, we identify four mechanisms that may potentially determine the ion selectivity in low-dimensional water.

The mechanism illustrated in Fig. \ref{fig2}a is independent of the confinement geometry and instead relies on the intrinsic ordered structure of water. When an ion enters a confined space, it must reorganize the surrounding water structure. If the relatively weak hydrogen bonds were unable to preserve this order, the ions would simply reorient the water molecules, forming successive hydration shells and diffusing further into the bulk. In low-dimensional water, however, a certain molecular ordering is maintained, meaning that an ion must break a certain number of hydrogen bonds when transitioning from the bulk into confinement.
The ability of an ion to enter the channel then is determined by the balance between the energy required to break these hydrogen bonds and the hydration energy gained. This provides a natural mechanism for ion selectivity, governed purely by the intrinsic properties of water. Such a phenomenon is commonly discussed \cite{ball2008water} in the context of protein \cite{hua2007nanoscale} and DNA folding \cite{cui2007DNAsolvent}, and, in theory, could be exploited for ion sieving below the nanoscale. However, in practical applications, the strength of hydrogen bonds may be insufficient to compete with the electric field of a bare ion. Ultimately, the effect is highly dependent on the stability of the water structure, which, as discussed in previous sections, can vary substantially.

The second ion selectivity mechanism is one of the most obvious and widely employed principles in ion sieving by artificial membranes. This form of selectivity arises purely from steric effects \cite{gopinadhan2019complete}. 
As illustrated in FIG. \ref{fig2}b, when the hydrated ion diameter is smaller than the confinement size ($d_h<d$) the ion is able to enter the channel along with the water flow. Conversely, if the ion's hydrated diameter exceeds the channel dimensions, it is generally expected to be excluded. However, as shown in FIG. \ref{fig2}c, this is not always the case.

Fig. 2c demonstrates that when $d_h>d$, an ion may still permeate the narrow channel, but this process necessitates at least partial dehydration. The associated energy cost $n_{dh}W_h/n_h$ corresponds to the dehydration penalty, which must be offset either by van der Waals interactions between the hydrated ion and the confining material or by an external electrostatic force driving ion transport. Specifically, the figure illustrates the partial dehydration of Na$^+$ with an initial hydration number of 
$n_h=6$,  a hydration energy of $W_h=4.2$ eV, and a hydrated diameter of $d_h=7.16$\AA \cite{zhu2019structure}
upon entering a confinement where $d<d_h$ resulting in a reduction of the hydration number by $n_{dh}=2$.

The energy barrier that must be overcome depends on both the properties of the confining material and the specific geometry of the channel. The dehydration penalty increases in narrower, lower-dimensional channels and for ions with higher hydration energies.
For graphene-based nanochannels, the barrier heights for different pore sizes are well established \cite{abraham2017tunable}. 
This steric-exclusion mechanism, when coupled with controlled dehydration effects \cite{razmjou2019design,goutham2023beyond}, represents one of the most advanced ion selectivity strategies, underpinning the exceptional functionality of natural ionic channels \cite{kumar2007highly}.

Beyond confinement geometry and material, functional groups at the channel entrance or within the channel itself can also influence ionic selectivity \cite{noskov2004control}. Depending on external conditions \cite{he2013bioinspired}, these groups may carry charge, modifying the surface charge of the channel walls and creating local Coulomb interactions that attract or repel specific ions \cite{zhao2013ion}. The resulting electric double layer can compete with steric confinement in regulating ion retention.
It is important to note that the dielectric permittivity of low-dimensional water is strongly reduced 
\cite{fumagalli2018anomalously,jalali2020out} enhancing Coulomb interactions within the electric double layer.
Certain materials (MXenes) have an exceptional ability to polarize water molecules giving rise 
to effectively negative permittivity \cite{sugahara2019negative,jalali2021hydration}.

Direct Coulomb interactions between an ion and a charged functional group at a binding site are much stronger than those with water molecules. This can lead to complete hydration shell disruption and even ion immobilization on the surface. Fig.  FIG. \ref{fig2}d  illustrates such an interaction between a positively charged ion and a negatively charged carboxyl group on graphene oxide. Unlike the mechanism in FIG. \ref{fig2}c, here ions with lower dehydration energy are more likely to be retained and filtered out. Selective ion separation can also be achieved by functionalizing nanochannel edges with different chemical groups.
Thus, to enhance the selectivity, CNTs have been functionalized using ligands such as biotin \cite{hinds2004aligned},
aliphatic groups \cite{majumder2005effect}, charged groups \cite{fornasiero2008ion} and zwitterions \cite{chan2013zwitterion}.

\section{Outlook}
\label{outro}

To conclude, we have offered a vision of low-dimenional water in the context of its intrinsic viscosity and structure.
A long-standing theoretical goal in water research is the development of a universal water model capable of accurately describing its behaviour across different conditions and environments reviewed in Table \ref{tab1}. This so-called `holy grail' of water theory remains elusive, as no single model can yet capture the complexities of water’s interactions, phase transitions, and anomalous properties in reduced dimensions.
A fundamental approach towards the model would be based on first-principles calculations which have so far proven prohibitively expensive and limited to a small number of molecules \cite{babin2012toward}.
Recently, one has been able to circumvent this difficulty by using a machine-learning model trained on density-functional theory energies and forces \cite{behler2016perspective}.
In this way one combines the accuracy of first-principles methods with the efficiency of simple force fields. The resulting force fields have the potential to lead to a water model that accurately reproduces all properties of water simultaneously and gives an accurate account of the interaction between water and the confining walls \cite{yu2023status}. 

A deeper understanding of water’s interactions at the quantum level is another pressing challenge \cite{min-visc}. The coupling between water molecules and the electron dynamics of confining surfaces could lead to quantum friction, a phenomenon that remains largely theoretical \cite{kavokine2022fluctuation}. Experimental investigations are needed to determine whether solid surfaces become polarized by water or electrolyte molecules and to what extent this influences transport properties. Additionally, harnessing electric fields to control water and ion flow is a major frontier in nanofluidics \cite{su2011control,fernandez2016electric,montenegro2021asymmetric}. Similarly, acoustic waves induced in confining walls could serve as a novel mechanism for manipulating dynamics of water not only at the microscale \cite{yeo2014surface} but in low-dimensional limits.

The phase behaviour of water at reduced dimensionality remains an unresolved issue, requiring experimental verification of its 1D and 2D phase diagrams presented in Fig. \ref{fig0}. Reproducing these exotic phases in the lab could provide insights into confined water's unique properties, such as enhanced transport and structural rearrangements. Simultaneously, achieving real-time measurements of ion concentrations in low-dimensional water with high spatial and temporal resolution would be transformative for both fundamental science and practical applications, such as biosensing and lab-on-a-chip technologies employing biological \cite{murata2000structural,sui2001structural} and biomimetic 1D \cite{siwy2017improving,shen2022fluorofoldamer} 
and 2D \cite{yang2024graphene,chen2025ultra,yang2022electro,bong2024graphene,yang2023nanofluidic,yang2023graphene,andreeva2021two} water nanochannels.

\bibliography{ice-vs-water.bib,2Dwater.bib,aqua.bib}

\section*{Acknowledgements}

This research is supported by the Singapore Ministry of Education Research Centre of Excellence award to the Institute for Functional Intelligent Materials
(I-FIM, Project No. EDUNC-33-18-279-V12).

\section*{Author contributions}

M.T. and K.S.N. developed the concept of the manuscript.
M.T. and D.V.A. collected data.
M.T. and F.M.P. wrote the manuscript. 

\section*{Competing interests}

The authors declare no competing interests.

\end{document}